\documentclass[aps,pre,twocolumn,groupaddress]{revtex4}

\usepackage{amsmath,makeidx,tabularx,float}
\usepackage[dvipdfm]{graphicx}
\usepackage{color}
\usepackage{ulem}

\begin{document}

\title{Reply to Chakrabarty et al.: Particles move even in ideal glasses}

\author{Misaki Ozawa}
\affiliation{Department of Physics, Nagoya
University, Nagoya 464-8602, Japan}

\author{Walter Kob}
\affiliation{Laboratoire Charles Coulomb, UMR 5221, 
Universit{\'e} of Montpellier and CNRS, F-34095 Montpellier, France}

\author{Atsushi Ikeda}
\affiliation{Fukui Institute for Fundamental Chemistry, Kyoto
University, Kyoto, 606-8103, Japan}

\author{Kunimasa Miyazaki}
\affiliation{Department of Physics, Nagoya
University, Nagoya, 464-8602, Japan}

%\contributor{Submitted to Proceedings of the National Academy of Sciences
%of the United States of America}

\maketitle

In their letter, Chakrabarty et al.~\cite{ckd}, point out
that their data on the relaxation dynamics are
inconsistent with the thermodynamic data
presented in our paper~\cite{okim}. 
They argue that from their results and the predictions of the
random first-order transition theory~\cite{cb_12} one
must conclude that our configurational entropy
$s_{\rm c}$ is “quantitatively not accurate.” 
In the following we will show that this conclusion
is not necessarily valid.

The main argument of Chakrabarty et al.~\cite{ckd}
(figure 1, Left, of Ref.~\cite{ckd}) is that the self
part of the intermediate scattering function
$F_s(k,t)$ decays to zero even in the glass phase
(defined by $s_{\rm c}=0$) and that hence this phase
cannot be nonergodic. 
Although this observation
is interesting, it does not imply the
mentioned conclusion. This can be seen by
considering the case of a simple crystal. In
a crystal the diffusion constant of a tagged
particle is nonzero (because a crystal at
nonzero temperatures will have a finite
concentration of defects) and hence $F_s(k,t)$ decays to zero at long times. From this one
can, however, not conclude that a crystal is
an ergodic system because its structure is
obviously independent of time (i.e., the collective correlation functions do not decay
to zero).

Thus, also in the case of the ideal glass
transition in the pinned system one has to
study not only the time dependence of the self
functions, but also the one of the collective
functions. This is done in FIG. 1, where we
compare the time dependence of these two
functions for different values of $c$ at the
temperature at which Chakrabarty et al.~\cite{ckd} have presented their data. 
The arrows indicate the static overlap function from
Fig. 2C of our article~\cite{okim}, that is, the
expected long time limit of the collective
function, showing the consistency of statics
and dynamics. These data also show
that with increasing $c$ the self and collective
functions decouple from each other
and that the relaxation time of the latter increases
rapidly, and thus shows a behavior
that is qualitatively consistent with the presence
of a nonergodic transition.

Regarding the accuracy of our configurational
entropy, we point out that in our paper~\cite{okim} we obtained the entropy by using a thermodynamic
integration of the potential
energy from the high temperature limit to
temperatures inside the ideal glass state. In a
completely independent calculation, we have
estimated the vibrational entropy for the
glass state via a harmonic approximation
that should give accurate results at lower
temperatures. Because inside the glass state
the two approaches give the same result, we
feel confident that the configurational entropy
has been obtained with good accuracy.

Finally, we mention that the concentration
for which Chakrabarty et al.~\cite{ckd} present their
results is close to or beyond the endpoint of
the glass transition line (Fig. 4 in our paper~\cite{okim}). Thus, unusual dynamics can be expected,
and it will be interesting to study this dynamics
in more detail.

\begin{figure}[h]
\begin{center}
\includegraphics[width=0.48\textwidth,clip]{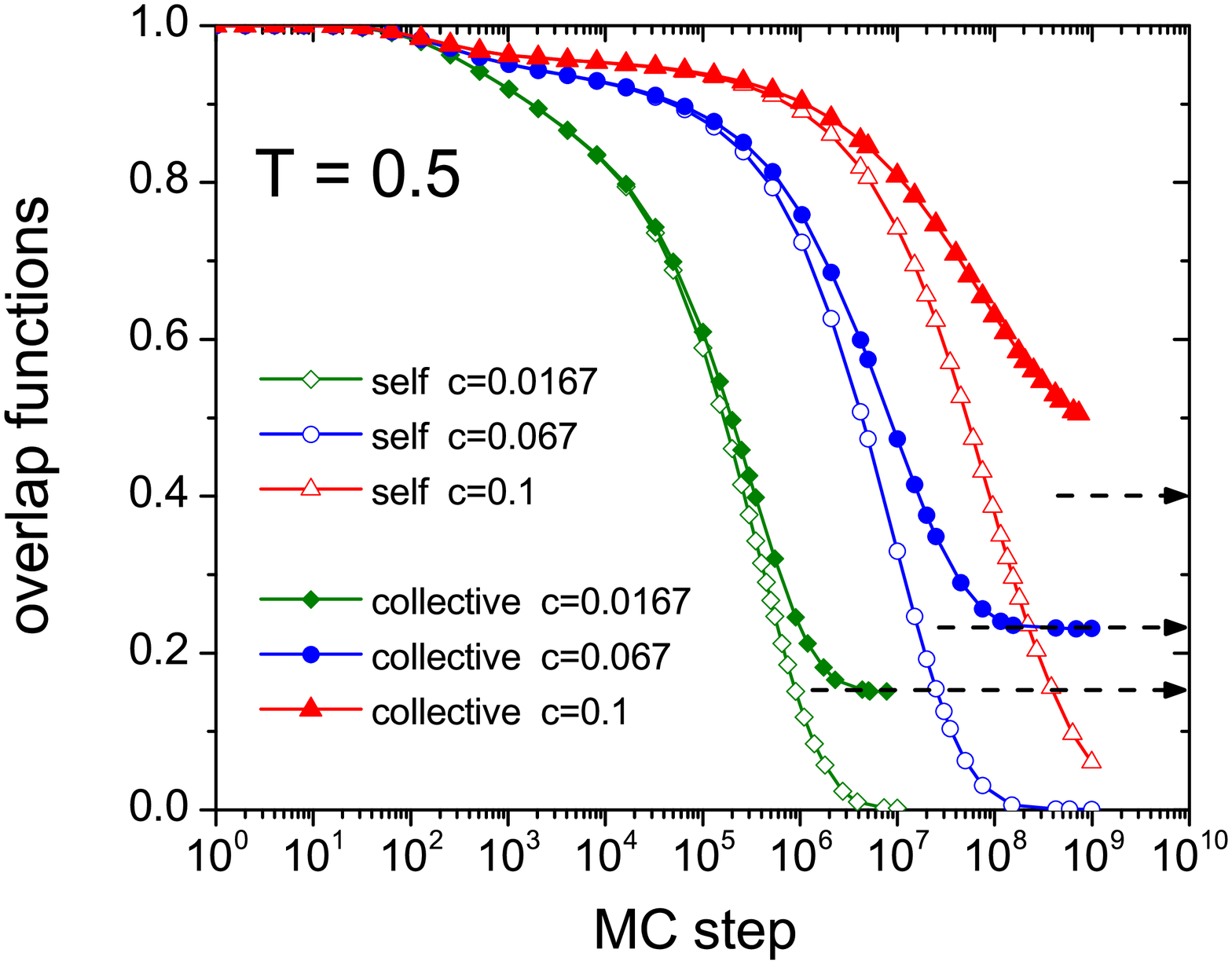}
\caption{
Time dependence of the self and collective overlap correlation functions (open and filled symbols, respectively) for pinning concentrations $c$ approaching the ideal glass
transition of a pinned system at $T=0.5$. 
The horizontal arrows indicate the value of the static overlap function from Fig. 2C of our paper~\cite{okim} and show that the static data are
consistent with the dynamic data at long times.
}
\end{center}
\label{fig1}
\end{figure}

\begin{acknowledgments}
We thank Chandan Dasgupta for helpful discussions.
MO acknowledge the financial support by Grant-in-Aid for JSPS Fellows (26.1878).
WK acknowledges the Institut Universitaire de France.
AI acknowledges JSPS KAKENHI No. 26887021. 
KM and MO acknowledge KAKENHI 
No. 24340098, % Kiban B
25103005, % Shingakujutu
25000002, % Tokubetsu suishin
and the JSPS Core-to-Core Program.
The simulations have been done in Research Center
for Computational Science, Okazaki, Japan, at the HPC@LR, and the CINES (grant c2014097308).
\end{acknowledgments}

\makeatletter
\renewcommand{\theequation}{
\thesection.\arabic{equation}}
\@addtoreset{equation}{section}
\makeatother

\end{document}